\documentstyle[12pt,epsfig]{article}
\date{}
\begin{document}

\newcommand{\mc}{\multicolumn}
\newcommand{\bce}{\begin{center}}
\newcommand{\ece}{\end{center}}
\newcommand{\beq}{\begin{equation}}
\newcommand{\eeq}{\end{equation}}
\newcommand{\bea}{\begin{eqnarray}}

\newcommand{\eea}{\end{eqnarray}}
\newcommand{\cont}{\nonumber\eea\bea}
\newcommand{\cl}[1]{\begin{center} {#1} \end{center}}
\newcommand{\ba}{\begin{array}}
\newcommand{\ea}{\end{array}}

\newcommand{\ab}{{\alpha\beta}}
\newcommand{\cd}{{\gamma\delta}}
\newcommand{\dc}{{\delta\gamma}}
\newcommand{\ac}{{\alpha\gamma}}
\newcommand{\bd}{{\beta\delta}}
\newcommand{\abc}{{\alpha\beta\gamma}}
\newcommand{\eps}{{\epsilon}}
\newcommand{\lam}{{\lambda}}
\newcommand{\mn}{{\mu\nu}}
\newcommand{\mpnp}{{\mu'\nu'}}
\newcommand{\Amuu}{{A_{\mu}}}
\newcommand{\Amuo}{{A^{\mu}}}
\newcommand{\Vmuu}{{V_{\mu}}}
\newcommand{\Vmuo}{{V^{\mu}}}
\newcommand{\Anuu}{{A_{\nu}}}
\newcommand{\Anuo}{{A^{\nu}}}
\newcommand{\Vnuu}{{V_{\nu}}}
\newcommand{\Vnuo}{{V^{\nu}}}
\newcommand{\Fmnu}{{F_{\mu\nu}}}
\newcommand{\Fmno}{{F^{\mu\nu}}}

\newcommand{\abcd}{{\alpha\beta\gamma\delta}}


\newcommand{\bsigma}{\mbox{\boldmath $\sigma$}}
\newcommand{\btau}{\mbox{\boldmath $\tau$}}
\newcommand{\brho}{\mbox{\boldmath $\rho$}}
\newcommand{\bpipi}{\mbox{\boldmath $\pi\pi$}}
\newcommand{\bss}{\bsigma\!\cdot\!\bsigma}
\newcommand{\btt}{\btau\!\cdot\!\btau}
\newcommand{\bnabla}{\mbox{\boldmath $\nabla$}}
\newcommand{\bphi}{\mbox{\boldmath $\tau$}}
\newcommand{\bvarphi}{\mbox{\boldmath $\rho$}}
\newcommand{\bDelta}{\mbox{\boldmath $\Delta$}}
\newcommand{\bpsi}{\mbox{\boldmath $\psi$}}
\newcommand{\bPsi}{\mbox{\boldmath $\Psi$}}
\newcommand{\bPhi}{\mbox{\boldmath $\Phi$}}
\newcommand{\bnab}{\mbox{\boldmath $\nabla$}}
\newcommand{\bpi}{\mbox{\boldmath $\pi$}}
\newcommand{\btheta}{\mbox{\boldmath $\theta$}}
\newcommand{\bkappa}{\mbox{\boldmath $\kappa$}}

\newcommand{\bA}{{\bf A}}
\newcommand{\bfe}{{\bf e}}
\newcommand{\bb}{{\bf b}}
\newcommand{\br}{{\bf r}}
\newcommand{\bj}{{\bf j}}
\newcommand{\bk}{{\bf k}}
\newcommand{\bl}{{\bf l}}
\newcommand{\bL}{{\bf L}}
\newcommand{\bM}{{\bf M}}
\newcommand{\bp}{{\bf p}}
\newcommand{\bq}{{\bf q}}
\newcommand{\bR}{{\bf R}}
\newcommand{\bs}{{\bf s}}
\newcommand{\bS}{{\bf S}}
\newcommand{\bT}{{\bf T}}
\newcommand{\bv}{{\bf v}}
\newcommand{\bV}{{\bf V}}
\newcommand{\bx}{{\bf x}}
\newcommand{\fph}{${\cal F}$}
\newcommand{\aph}{${\cal A}$}
\newcommand{\dph}{${\cal D}$}
\newcommand{\fpi}{f_\pi}
\newcommand{\mpi}{m_\pi}
\newcommand{\Tr}{{\mbox{\rm Tr}}}
\def\Qb{\overline{Q}}
\newcommand{\delu}{\partial_{\mu}}
\newcommand{\delo}{\partial^{\mu}}
%
%
\newcommand{\up}{\!\uparrow}
\newcommand{\upup}{\uparrow\uparrow}
\newcommand{\updo}{\uparrow\downarrow}
\newcommand{\uu}{$\uparrow\uparrow$}
\newcommand{\ud}{$\uparrow\downarrow$}
\newcommand{\auu}{$a^{\uparrow\uparrow}$}
\newcommand{\aud}{$a^{\uparrow\downarrow}$}
\newcommand{\pu}{p\!\uparrow}

\newcommand{\qp}{quasiparticle}
\newcommand{\sa}{scattering amplitude}
\newcommand{\ph}{particle-hole}
\newcommand{\qcd}{{\it QCD}}
\newcommand{\integ}{\int\!d}
\newcommand{\ie}{{\sl i.e.~}}
\newcommand{\etal}{{\sl et al.~}}
\newcommand{\etc}{{\sl etc.~}}
\newcommand{\rhs}{{\sl rhs~}}
\newcommand{\lhs}{{\sl lhs~}}
\newcommand{\eg}{{\sl e.g.~}}
\newcommand{\ef}{\epsilon_F}
\newcommand{\sigt}{d^2\sigma/d\Omega dE}
\newcommand{\sige}{{d^2\sigma\over d\Omega dE}}
\newcommand{\rpaeq}{\beq
\left ( \begin{array}{cc}
A&B\\
-B^*&-A^*\end{array}\right )
\left ( \begin{array}{c}
X^{(\kappa})\\Y^{(\kappa)}\end{array}\right )=E_\kappa
\left ( \begin{array}{c}
X^{(\kappa})\\Y^{(\kappa)}\end{array}\right )
\eeq}
\newcommand{\ket}[1]{| {#1} \rangle}
\newcommand{\bra}[1]{\langle {#1} |}
\newcommand{\ave}[1]{\langle {#1} \rangle}

\newcommand{\singlespace}{
    \renewcommand{\baselinestretch}{1}\large\normalsize}
\newcommand{\doublespace}{
    \renewcommand{\baselinestretch}{1.6}\large\normalsize}
\newcommand{\bftau}{\mbox{\boldmath $\tau$}}
\newcommand{\bfalpha}{\mbox{\boldmath $\alpha$}}
\newcommand{\bfgamma}{\mbox{\boldmath $\gamma$}}
\newcommand{\bfxi}{\mbox{\boldmath $\xi$}}
\newcommand{\bfbeta}{\mbox{\boldmath $\beta$}}
\newcommand{\bfeta}{\mbox{\boldmath $\eta$}}
\newcommand{\bfpi}{\mbox{\boldmath $\pi$}}
\newcommand{\bfphi}{\mbox{\boldmath $\phi$}}
\newcommand{\bfR}{\mbox{\boldmath ${\cal R}$}}
\newcommand{\bfL}{\mbox{\boldmath ${\cal L}$}}
\newcommand{\bfM}{\mbox{\boldmath ${\cal M}$}}
\def\dblint{\mathop{\rlap{\hbox{$\displaystyle\!\int\!\!\!\!\!\int$}}
    \hbox{$\bigcirc$}}}
\def\ut#1{$\underline{\smash{\vphantom{y}\hbox{#1}}}$}

\def\Pom{{\bf I\!P}}
\def\lsim{\mathrel{\rlap{\lower4pt\hbox{\hskip1pt$\sim$}}
    \raise1pt\hbox{$<$}}}         
\def\gsim{\mathrel{\rlap{\lower4pt\hbox{\hskip1pt$\sim$}}
    \raise1pt\hbox{$>$}}}         
\def\beq{\begin{equation}}
\def\eeq{\end{equation}}
\def\bea{\begin{eqnarray}}
\def\eea{\end{eqnarray}}

\begin{center}
{\Large\bf  Saturation of Nuclear Partons: the Fermi Statistics or
Nuclear Opacity?  }\\ \vspace{1cm}
 { \bf N.N. Nikolaev$^{a,b)}$,
W. Sch\"afer$^{c)}$, B.G. Zakharov$^{b)}$, V.R.
Zoller$^{d)}$\medskip\\  }

$^{a)}$ Institut f. Kernphysik, Forschungszentrum J\"ulich, D-52425 J\"ulich, Germany\\
$^{b)}$ L.D.Landau Institute for Theoretical Physics, Chernogolovka, Russia\\
$^{c)}$ NORDITA, Blegdamsvej 17, DK-2100 Copenhagen \O, Denmark\\
$^{d)}$ Institute for Theoretical and Experimental Physics, Moscow, Russia\\
E-mail: N.Nikolaev$@$fz-juelich.de\vspace{1cm} \\

{\bf Abstract\\    }

\end{center}
{\small We derive the two-plateau momentum distribution of final
state (FS) quarks produced in  deep inelastic scattering (DIS) off
nuclei in the saturation regime. The diffractive plateau which
dominates for small $\bp$ measures precisely the momentum
distribution of quarks in the beam photon, the r\^ole of the
nucleus is simply to provide an opacity. The plateau for truly
inelastic DIS exhibits a substantial nuclear broadening of the FS
momentum distribution. We discuss the relationship between the FS
quark densities and the properly defined initial state (IS)
nuclear quark densities.The Weizs\"acker-Williams glue of a
nucleus exhibits a substantial nuclear dilution, still soft IS
nuclear sea saturates because of the anti-collinear splitting of
gluons into sea quarks. \medskip\\}


The interpretation of nuclear opacity in terms of a fusion
and saturation of nuclear partons has been introduced in
1975 \cite{NZfusion} way before the QCD parton model:
 the
Lorentz contraction of relativistic nuclei entails a spatial
overlap of partons with $x \lsim x_{A} \approx 1/R_A m_N$ from
different nucleons and the fusion of overlapping partons results
in the saturation of parton densities per unit area in the impact
parameter space. The pQCD link between nuclear opacity and
saturation has been considered  by Mueller \cite{Mueller1} and the
pQCD discussion of fusion of nuclear gluons has been revived by
McLerran et al. \cite{McLerran}.

The common wisdom is that in DIS the FS interaction effects can be
neglected and the observed momentum distribution of struck partons
in the FS coincides with the IS density of partons in the probed
hadron. Based on the consistent treatment of intranuclear
distortions, we derive the two-plateau spectrum of FS quarks. We
find a substantial nuclear broadening of inclusive FS spectra and
demonstrate that despite this broadening the FS sea parton density
exactly equals the IS sea parton density calculated in terms of
the WW glue of the nucleus as defined according to \cite{NSS}. We
pay a special attention to an important point that diffractive DIS
in which the target nucleus does not break and is retained in the
ground state, makes precisely 50 per cent of the total DIS events
\cite{NZZdiffr}. We point out that the saturated diffractive
plateau measures precisely the momentum distribution of
(anti-)quarks in the $q\bar{q}$ Fock state of the photon. In
contrast to DIS off nuclei, the fraction of DIS off free nucleons
which is diffractive is negligibly small \cite{GNZ}, $\eta_D \lsim
$ 6-10 \%, and there is little room for genuine saturation effects
even at HERA. We show how the anti-collinear splitting of WW
gluons into sea quarks gives rise to nuclear saturation of the sea
despite a substantial nuclear dilution of the WW glue.

We base our analysis on the color dipole formulation of DIS
\cite{NZZdiffr,NZ91,NZ92,NZ94,NZZlett} and illustrate our ideas on
an example of DIS at $x\sim x_A \ll 1$ which is dominated by
interactions of $q\bar{q}$ states of the photon. The total cross
section for interaction of the color dipole $\br$ with the target
nucleon equals \bea \sigma(r)= \alpha_S(r) \sigma_0\int d^2\bkappa
f(\bkappa )\left[1 -\exp(i\bkappa \br )\right]\, , \label{eq:1}
\eea where $f(\bkappa )$ is related to the unintegrated glue of
the target nucleon by \bea f(\bkappa ) = {4\pi \over
N_c\sigma_0}\cdot {1\over \kappa^4} \cdot {\partial G \over
\partial\log\kappa^2}\, , ~~~~\int d^2\bkappa  f(\bkappa )=1\, .
\label{eq:2} \eea
 For DIS off a free nucleon target, see figs. 1a-1d, the momentum
spectrum of the FS
quark  prior the hadronization,
\bea
{d\sigma_N \over d^2\bp dz} =
{\sigma_0\over 2}\cdot { \alpha_S(\bp^2) \over (2\pi)^2}
 \int d^2\bkappa f(\bkappa )
\left|\langle \gamma^*|\bp\rangle - \langle \gamma^*|\bp-\bkappa \rangle\right|^2
\label{eq:3}
\eea
where $\bp$ is the transverse momentum, and $z$ the Feynman variable,
coincides, upon the $z$ integration, with
the conventional IS unintegrated $\bp$
distribution of partons in the target.
Notice that the target nucleon is color-excited and
there is no rapidity gap in the FS.

\begin{figure}[t]
\begin{center}
\epsfig{file=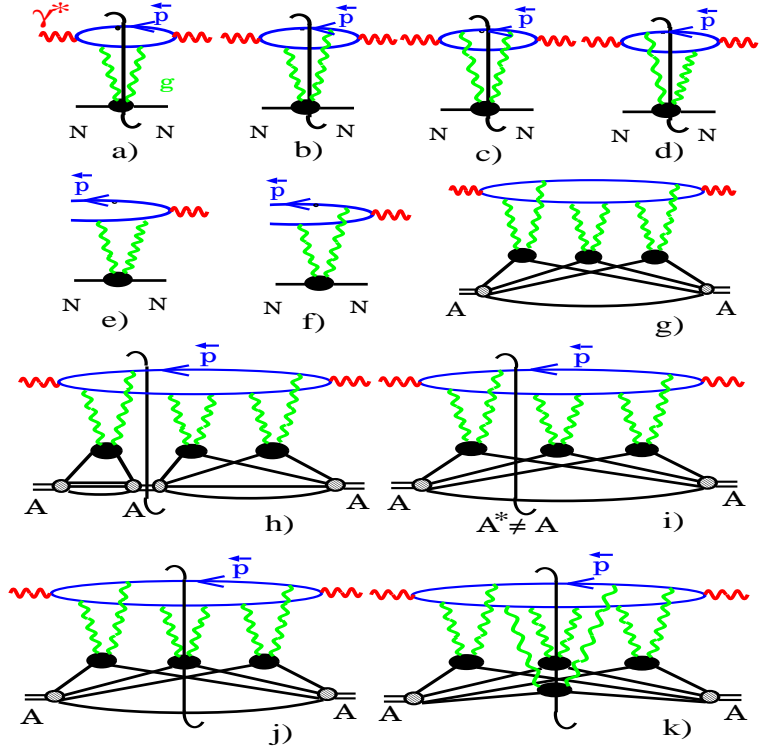, height=10.0cm, width = 10.0cm}
\end{center}
\caption{The pQCD diagrams for inclusive (a-d) and
diffractive
(e,f) DIS off protons and nuclei
(g-k). Diagrams (a-d) show the unitarity cuts with color
excitation of the target nucleon, (g) - a generic multiple
scattering diagram for Compton scattering off nucleus, (h) - the
unitarity cut for a coherent diffractive DIS, (i) - the unitarity
cut for quasielastic diffractive DIS with excitation of the
nucleus $A^*$, (j,k) - the unitarity cuts for truly inelastic DIS
with single and multiple color excitation of nucleons of the
nucleus. }
\end{figure}

In  DIS off nuclei one must distinguish the three principal
processes with distinct unitarity cuts of the forward Compton
amplitude (fig.~1g): the coherent diffraction dissociation {\sl
(D)} of the photon (fig.~1h), quasielastic diffraction
dissociation {\sl (qel)} followed by excitation and breakup of the
target nucleus  (fig.~1i) - in both of them there is no color flow
between the photon debris and the nucleus - , and the truly
inelastic {\sl (in)} DIS with color excitation of nucleons of the
target nucleus (fig.~1j,k). A useful guidance for isolation of
different
 processes
comes from the coupled-channel formalism presented in \cite{NNNJETP}.

We work in the conventional approximation of two t-channel gluons
in DIS off free nucleons, figs. 1a-1d, i.e., neglecting the effect
of diffractive DIS (figs. 1e,f) on the total cross section on free
nucleons, $\eta_D \ll 1$. Then the S-matrix of the quark-nucleon
scattering must be computed to the second order in the QCD eikonal
$\delta(\bb)$, and the S-matrix for the color dipole-nucleon
scattering takes the form \beq S(\bb_+,\bb_-)=1 + 2i
[\Delta(\bb_+)-\Delta(\bb_-)] -
2\langle[\delta(\bb_+)-\delta(\bb_-)]^2\rangle_0 \label{eq:4} \eeq
Here $\langle...\rangle_0$ indicates that we must take only the
color singlet component of the two-gluon exchange, and we
introduced special notation $\Delta(\bb)=\delta(\bb)$ for the
color-exchange component of the S-matrix. The color dipole cross
section equals $\sigma(\br) = 2 \int d^2\bb \left\{1- \langle
S(\bb+\br,\bb)\rangle_0\right\}$, which relates the QCD eikonal to
the gluon structure function of the nucleon, \beq \int d^2\bb
\langle \delta(\bb+\br)\delta(\bb)\rangle_0= {1\over 8}
\alpha_S(\br)\sigma_0 \int d^2\bkappa f(\bkappa)\exp[i\bkappa\br]
\label{eq:5} \eeq If in the nuclear S-matrix $
S_A(\{\Delta\},\{\delta\};\bb_+,\bb_-) = \prod _{j=1}^A
S(\bb_+-\bb_j,\bb_- - \bb_j) $ one puts $\Delta
(\bb_{\pm}-\bb_j)\equiv 0$, then it would describe pure
diffraction without color excitations in a nucleus. Then, with
standard reference to closure \cite{NNNJETP}, the momentum
spectrum of observed FS quarks for truly inelastic DIS with color
excitations in the nucleus can readily be isolated: \bea
{d\sigma_{in} \over dz d^2\bp_+ d^2\bp_-} &=&
{1\over (2\pi)^4} \int d^2 \bb_+' d^2\bb_-' d^2\bb_+ d^2\bb_- \nonumber\\
&\times& \exp[-i\bp_+(\bb_+ -\bb_+')-i\bp_-(\bb_- -
\bb_-')]\Psi^*(\bb_+' -\bb_-')
\Psi(\bb_+ -\bb_-)\nonumber\\
&\times& \left\{ \langle
A|S_A^*(\{\Delta\},\{\delta\};\bb_+',\bb_-')
S_A(\{\Delta\},\{\delta\};\bb_+,\bb_-)|A\rangle \right. \nonumber\\
&& - \left. \langle A|S_A^*(0,\{\delta\};\bb_+',\bb_-')
S_A(0,\{\delta\};\bb_+,\bb_-)|A\rangle \right\} \label{eq:6} \eea
Here $\Psi$ is the $q \bar{q}$--Fock state wave function of the
virtual photon, and we suppressed its dependence on $z$. The FS
spectra in the coherent and quasielastic diffraction are obtained
by the substitutions of the expression in the curly braces by \bea
\left\{...\right\}_{D}&=&\langle
A|1-S_A^*(0,\{\delta\};\bb_+',\bb_-')|A\rangle \langle
A|1-S_A(0,\{\delta\};\bb_+,\bb_-)|A\rangle \label{eq:7} \\
\left\{...\right\}_{qel}&=&\left\{ \langle
A|S_A^*(0,\{\delta\};\bb_+',\bb_-')
S_A(0,\{\delta\};\bb_+,\bb_-)|A\rangle \right. \nonumber\\
&&-\left. \langle A|S_A^*(0,\{\delta\};\bb_+',\bb_-')|A\rangle
\langle A|S_A(0,\{\delta\};\bb_+,\bb_-)|A\rangle \right\} \, .
\label{eq:8} \eea

The size of color dipoles can be neglected compared to the radius
of heavy nuclei. In the standard approximation of a dilute gas
nucleus only the color-singlet terms $\propto \langle
\Delta(\bb_1)\Delta(\bb_1)\rangle_0$ would appear in (\ref{eq:6}),
and we find \bea \langle
A|S_A^*(\{\Delta\},\{\delta\};\bb_+',\bb_-')
S_A(\{\Delta\},\{\delta\};\bb_+,\bb_-)|A\rangle = \exp\{- {1\over
2}T(\bb)\left[
    \Sigma(\bb_+' - \bb_+) \right. \nonumber \\
    + \Sigma(\bb_-' - \bb_-)
    \left.
    - \Sigma(\bb_+' - \bb_-)- \Sigma(\bb_+ - \bb_-') +\sigma(\bb_+' -
\bb_-')+ \sigma(\bb_+ - \bb_-)\right]\}\, , \label{eq:9} \eea were
$T(\bb)=\int dz n_{A}(z, \bb)$ is the optical thickness of a
nucleus at an impact parameter $\bb \approx \bb_{\pm},\bb_{\pm}'$.
Here $\Sigma(\br)=\sigma(\br)$, we use the capital letter just to
indicate that it originates from the color excitation processes,
whereas the last two terms in the exponent of (\ref{eq:9})
describe intranuclear attenuation of the $q\bar{q}$ pair due to
color-singlet exchanges. The diffractive S-matrix elements
entering (\ref{eq:7}), (\ref{eq:8}) are readily obtained from
(\ref{eq:9}). For a heavy nucleus the quasielastic diffraction is
a surface phenomenon and can be neglected for all the practical
purposes, see \cite{NZZdiffr}; it vanishes to the considered
leading order (\ref{eq:4}). If we are interested only in the
single particle spectrum and integrate over $\bp_-$, then
$\bb_-'=\bb_-$ and the diffractive attenuation terms in the
exponent of (\ref{eq:9}) would exactly cancel the two terms from
multiple color excitation processes in the last line of
(\ref{eq:9}), and \bea {d \sigma_{in}\over d^2\bp dz }   &=&  {1
\over (2\pi)^2}\int d^2\bb
 \int d^2\br' d^2\br
\exp[i\bp(\br'-\br)]\Psi^*(\br')\Psi(\br)\nonumber\\
&\times& \left\{\exp[-{1\over 2}\sigma(\br-\br')T(\bb)]-
\exp[-{1\over 2}[\sigma(\br)+\sigma(\br')]T(\bb)]\right\}\, .
\label{eq:10} \eea Evidently, the dependence of nuclear
attenuation factors on $\br,\br'$ shall distort strongly the
observed momentum distribution of quarks.

In the further interpretation of these results in terms of the
parton model  we resort to the NSS
representation \cite{NSS}
\bea
\Gamma_A(\bb,\br)= 1-\exp\left[-{1\over 2}\sigma(r)T(\bb)\right]
=
 \int d^2\bkappa  \phi_{WW}(\bkappa )[1-\exp(i
\bkappa \br) ] \, .
\label{eq:11}
\eea
Driven by a  close analogy to (\ref{eq:1}),(\ref{eq:2})
in terms of $f(\bkappa)$,  we interpret
\beq
 \phi_{WW}(\bkappa ) =
\sum_{j=1}^{\infty} \nu_A^j(\bb) \cdot {1 \over j!}
f^{(j)}(\bkappa ) \exp\left[-\nu_A(\bb)\right] \label{eq:12} \eeq
as the  Weizs\"acker-Williams (WW) unintegrated glue of a nucleus
per unit area in the impact parameter plane. It is normalized as
\beq \int d^2\bkappa \phi_{WW}(\bkappa )= 1-\exp[-\nu_A(\bb)]\,.
\label{eq:13} \eeq Here $$\nu_A(\bb)= {1 \over
2}\alpha_S(r)\sigma_0T(\bb)$$ defines the nuclear opacity and the
$j$-fold convolutions \beq f^{(j)}(\bkappa )= \int \prod_{i=1}^j
d^2\bkappa _{i} f(\bkappa _{i}) \delta(\bkappa -\sum_{i=1}^j
\bkappa _i) \label{eq:14} \eeq describe the contribution to the
diffractive amplitudes from $j$ split pomerons \cite{NSS}.

A discussion of the nuclear antishadowing property of the hard WW
glue is found in \cite{NSS}. A somewhat involved analysis of the
properties of the convolutions (\ref{eq:14}) in the soft region
shows that they develop a plateau-like behaviour with the width of
the plateau which expands $\propto j$. The gross features of the
WW nuclear glue in the soft region are well reproduced by \bea
\phi_{WW}(\bkappa) \approx  {1\over \pi}  {Q_A^2 \over (\bkappa^2
+Q_{A}^2)^2}\, , \label{eq:15} \eea where the saturation scale $
Q_A^2 =  \nu_A(\bb)  Q_0^2 \propto A^{1/3}\, . $ The soft
parameters $Q_0^2$ and $\sigma_0 $ are related to the integrated
glue of the proton in the soft region,
$$
Q_{0}^2\sigma_0 \sim {2\pi^2 \over N_c} G_{soft}\,, ~~~G_{soft}\sim 1\,.
$$
Notice the nuclear dilution of soft WW glue, $\phi_{WW}(\bkappa) \propto 1/Q_A^2 \propto
A^{-1/3}$.

On the one hand, making use of the NSS representation, the total
nuclear photoabsorption cross section can be cast in the form \beq
\sigma_{A}  = \int d^2\bb\int dz \int {d^2\bp\over (2\pi)^2} \int
d^2\bkappa\phi_{WW}(\bkappa) \left|\langle \gamma^* |\bp\rangle -
\langle \gamma^* |\bp-\bkappa\rangle \right|^2 \label{eq:16} \eeq
which has a profound  semblance to (\ref{eq:3}) and one is tempted
to take the differential form of (\ref{eq:16}) as a definition of
the IS sea quark density in a nucleus: \bea {d\bar{q}_{IS} \over
d^2\bb d^2\bp} = {1\over 2}\cdot{Q^2 \over 4\pi^2 \alpha_{em}}
\cdot{d\sigma_A \over d^2\bb d^2\bp}\, . \label{eq:17} \eea In
terms of the WW nuclear glue, all intranuclear multiple-scattering
diagrams of fig.~1g sum up to precisely the same four diagrams
fig.~1a-1d as in DIS off free nucleons. Furthermore, one can argue
that the small-$x$ evolution of the so-defined IS nuclear sea is
similar to that for a free nucleon sea. Although $\bp$ emerges
here just as a formal Fourier parameter, we shall demonstrate that
it can be identified with the momentum of the observed final state
antiquark.

On the other hand, making use of the NSS representation, after
some algebra one finds \bea \exp[-{1\over
2}\sigma(\br-\br')T(\bb)]- \exp[-{1\over
2}[\sigma(\br)+\sigma(\br')]T(\bb)] = \int d^2\bkappa
\phi_{WW}(\bkappa) \nonumber \\
 \{(\exp[i\bkappa(\br-\br')]-1) +
(1-\exp[i\bkappa\br])+(1-\exp[i\bkappa\br'])\}\nonumber\\ - \int
d^2\bkappa \phi_{WW}(\bkappa)(1-\exp[i\bkappa\br]) \int
d^2\bkappa' \phi_{WW}(\bkappa')(1-\exp[i\bkappa'\br'])~~~
\label{eq:18} \eea
 \bea {d \sigma_{in}\over d^2\bb d^2\bp dz } &=&
{1 \over (2\pi)^2}\left\{ \int  d^2\bkappa
\phi_{WW}(\bkappa)\left|\langle \gamma^* |\bp\rangle -
\langle \gamma^* |\bp-\bkappa\rangle \right|^2 \right. \nonumber\\
&& - \left. \left|\int d^2\bkappa\phi_{WW}(\bkappa) (\langle
\gamma^* |\bp\rangle - \langle \gamma^*
|\bp-\bkappa\rangle )\right|^2\right\} \label{eq:19} \\
{d \sigma_{D}\over d^2\bb d^2\bp dz }   &=&  {1 \over (2\pi)^2}
\left|\int d^2\bkappa\phi_{WW}(\bkappa) (\langle \gamma^*
|\bp\rangle - \langle \gamma^* |\bp-\bkappa\rangle)\right|^2 \, .
\label{eq:20} \eea As far as diffraction is concerned, the analogy
between (\ref{eq:20}) and its counterpart for free nucleons
\cite{NSS,NZ92,NZsplit}, and nuclear WW glue $\phi_{WW}(\bkappa)$
and $f(\bkappa)$ thereof, is complete. Putting the inelastic and
diffractive components of the FS quark spectrum together, we
evidently find the FS parton density which exactly coincides with
the IS parton density (\ref{eq:17}) such that $\bp$ is indeed the
transverse momentum of the FS sea quark. The interpretation of
this finding is not trivial, though.

Consider first the domain of $\bp^2 \lsim Q^2 \lsim Q_A^2$ such
that the nucleus is opaque for all color dipoles in the photon.
Hereafter we assume that the saturation scale $Q_A^2$ is so large
that $\bp^2,Q^2$ are in the pQCD domain and neglect the quark
masses. In this regime the nuclear counterparts of the crossing
diagrams of figs. 1b,d,f  can be neglected. Then, in the
classification of \cite{NSS}, diffraction will be dominated by the
contribution from the Landau-Pomeranchuk diagram of fig.~1e with
the result \bea \left. {d\bar{q}_{FS} \over d^2\bb
d^2\bp}\right|_D &=& {1\over 2}\cdot{Q^2 \over 4\pi^2 \alpha_{em}}
\cdot{d\sigma_D \over d^2\bb d^2\bp} \nonumber\\
&\approx& {1\over 2}\cdot{Q^2 \over 4\pi^2 \alpha_{em}} \cdot\int
dz \left| \int d^2\bkappa\phi_{WW}(\bkappa) \right|^2
\left|\langle \gamma^* |\bp\rangle\right|^2 \approx {N_c \over
4\pi^4}\, . \label{eq:21} \eea Up to now we specified neither the
wave function of the photon nor the spin  nor the color
representation of charged partons, only the last result in
(\ref{eq:17}) makes explicit use of the conventional spin-${1\over
2}$ partons. We also used the normalization (\ref{eq:13}).
Remarkably, diffractive DIS measures the momentum distribution of
quarks and antiquarks in the $q\bar{q}$ Fock state of the photon.
We emphasize that this result, typical of the Landau-Pomeranchuk
mechanism, is a completely generic one and would hold for any beam
particle such that its coupling to colored partons is weak. In
contrast to diffraction off free nucleons
\cite{NZ92,NZsplit,GNZcharm}, diffraction off opaque nuclei is
dominated by the anti-collinear splitting of hard gluons into soft
sea quarks, $\bkappa^2 \gg \bp^2$. Precisely for this reason one
finds the saturated FS quark density, because the nuclear dilution
of the WW glue is compensated for by the expanding plateau. The
result (\ref{eq:21}) has no counterpart in DIS off free nucleons
because diffractive DIS off free nucleons is negigibly small even
at HERA, $\eta_D \lsim $ 6-10 \%.

The related analysis of the FS quark density for truly inelastic
DIS in the  same domain of $\bp^2 \lsim Q^2 \lsim Q_A^2$ gives
\bea \left.{d\bar{q}_{FS} \over d^2\bb d^2\bp}\right|_{in} &=&
{1\over 2}\cdot{Q^2 \over 4\pi^2 \alpha_{em}} \cdot\int dz \int
d^2\bkappa \phi_{WW}(\bkappa)
\left|\langle \gamma^* |\bp-\bkappa\rangle \right|^2 \nonumber\\
 &=&
{Q^2 \over 8\pi^2 \alpha_{em}}\phi_{WW}(0)
\int^{Q^2} d^2\bkappa \int dz \left|\langle \gamma^* |\bkappa\rangle \right|^2
= {N_c \over 4\pi^4}\cdot {Q^2 \over Q_A^2}\cdot\theta(Q_A^2-\bp^2)\, .
\label{eq:22}
\eea
It describes final states with color excitation of a nucleus,
but as a function of the photon wave
function and nuclear WW gluon distribution it is completely
different from (\ref{eq:3}) for free nucleons. The $\theta$-function simply
indicates that the plateau for inelastic DIS extends up to $\bp^2
\lsim Q^2_A$.
For $Q^2 \ll Q_A^2$ the inelastic plateau
contributes little to the transverse momentum distribution of
soft quarks, $\bp^2 \lsim Q^2$, but the inelastic plateau extends way beyond
$Q^2$ and its integral contribution to the spectrum of FS
quarks is exactly equal to that from diffractive DIS. Such a  two-plateau
structure of the FS quark spectrum is a new finding and has not been
considered before.

Now notice, that  in the opacity regime the diffractive FS parton
density coincides with the contribution $\propto |\langle
\gamma^*|\bp\rangle|^2$ to the IS sea parton density from the
spectator diagram 1a, whereas the FS parton density for truly
inelastic DIS coincides with the contribution to IS sea partons
from the diagram of fig.~1c. The contribution from the crossing
diagrams 1b,d is negigibly small.

Our results (\ref{eq:21}) and (\ref{eq:22}), especially nuclear broadening
and unusually strong $Q^2$ dependence of the FS/IS parton density from
truly inelastic DIS, demonstrate clearly a distinction between diffractive
and inelastic DIS. Our considerations can readily be extended to the
spectrum of soft quarks, $\bp^2 \lsim Q_A^2$, in hard photons, $Q^2 \gsim Q_{A}^2$.
In this case the result (\ref{eq:21}) for diffractive DIS is retained,
whereas in the numerator of the result (\ref{eq:22}) for truly inelastic
DIS one must substitute $Q^2 \to Q_{A}^2$, so that in this case
$dq_{FS}|_{D} \approx dq_{FS}|_{in}$ and $dq_{IS} \approx
2dq_{FS}|_{D}$. The evolution
of soft nuclear sea, $\bp^2 \lsim Q_{A}^2$, is entirely driven by
an anti-collinear splitting of the NSS-defined WW nuclear glue into the sea
partons.

The early discussion of the FS quark density in the saturation
regime is due to Mueller \cite{Mueller}. Mueller focused on $Q^2
\gg Q_A^2$ and discussed  neither a distinction between
diffractive and truly inelastic DIS nor a $Q^2$ dependence and
broadening (\ref{eq:15}) for truly inelastic DIS at $Q^2 \lsim
Q_A^2$.

We come to a summary. We reported a derivation of the FS parton
density. Our result (\ref{eq:9}) summarizes in an elegant way
intranuclear distortions due to multiple diffractive rescatterings
and color excitations of the target nucleus. In conjunction with
the NSS definition of the WW glue of the nucleus, eqs.~
(\ref{eq:11}), (\ref{eq:12}), it gives an explicit form of the FS
parton densities. The two-plateau FS quark density with the strong
$Q^2$ dependence of the plateau for truly inelastic DIS has not
been discussed before. A comparison with the IS nuclear parton
densities which evolve from the NSS-defined WW nuclear glue shows
an exact equality of the FS and IS parton densities. The
plateau-like saturated nuclear quark density is suggestive of the
Fermi statistics, but our principal point that for any projectile
which interacts weakly with colored partons the saturated density
measures the momentum distribution in the $q\bar{q} , gg,...$ Fock
state of the projectile disproves the Fermi-statistics
interpretation. The spin and color multiplet of colored partons
the photon couples to is completely irrelevant, what only counts
is an opacity of heavy nuclei. The anti-collinear splitting of WW
nuclear glue into soft sea partons is a noteworthy feature of the
both diffractive DIS and IS sea parton distributions. The
emergence of a saturated density of IS sea partons from the
nuclear-diluted WW glue is due to the nuclear broadening of the
plateau (\ref{eq:15}). Because the predominance of diffraction is
a very special feature of DIS \cite{NZZdiffr}, one must be careful
with applying the IS parton densities to, for instance, nuclear
collisions, in which diffraction wouldn't be of any significance.

One can go one step further and consider interactions with the
opaque nucleus of the $q\bar{q}g$ Fock states of the photon.
Then the above analysis can be extended to $x \ll x_A$  and
the issue of the $x$-dependence of the saturation scale
$Q_A^2$ can be addressed following the discussion in \cite{NZ94}.
We only mention here that as far as diffraction and IS parton
densities are concerned, the NSS-defined WW glue
remains a useful concept and the close correspondence between
$\phi_{WW}(\bkappa)$ for the nucleus and $f(\bkappa)$ for the nucleon is
retained.

This work has been partly supported by the INTAS grants 97-30494
\& 00-00366 and the DFG grant 436RUS17/119/01, and by DAAD and
Nordita.

\end{document}